\documentclass[aps,prx,reprint,superscriptaddress,amsmath,showpacs,amsfonts,amssymb]{revtex4-1}
\usepackage{graphicx}
\usepackage{bm}
\usepackage{color}
\usepackage{MnSymbol}
\usepackage{enumerate}
\usepackage{bbold}
\usepackage{lipsum}
\usepackage{array}

\def\ra{\rangle}

\def\be{\begin{equation}}
\def\ee{\end{equation}}

\newcommand{\ket}[1]{\vert#1\ra}                 
\newcommand{\Tr}[1]{\text{Tr}(#1)}               

\begin{document}

\title{Quantum trajectories and their statistics for remotely entangled quantum bits}

\author{Areeya Chantasri}
\affiliation{Department of Physics and Astronomy, University of Rochester, Rochester, NY 14627, USA}
\affiliation{Center for Coherence and Quantum Optics, University of Rochester, Rochester, NY 14627, USA}
\author{Mollie E. Kimchi-Schwartz}
\affiliation{Quantum Nanoelectronics Laboratory, Department of Physics, University of California, Berkeley, California 94720, USA}
\author{Nicolas Roch}
\affiliation{Universit{\'e} Grenoble Alpes, Institut NEEL, F-38000 Grenoble, France and CNRS, Institut NEEL, F-38000 Grenoble, France}
\author{Irfan Siddiqi}
\affiliation{Quantum Nanoelectronics Laboratory, Department of Physics, University of California, Berkeley, California 94720, USA}
\author{Andrew N. Jordan}
\affiliation{Department of Physics and Astronomy, University of Rochester, Rochester, NY 14627, USA}
\affiliation{Center for Coherence and Quantum Optics, University of Rochester, Rochester, NY 14627, USA}
\affiliation{Institute for Quantum Studies, Chapman University, 1 University Drive,
Orange, CA 92866, USA}

\date{\today}

\begin{abstract}
We experimentally and theoretically investigate the quantum trajectories of jointly monitored transmon qubits embedded in spatially separated microwave cavities. Using nearly quantum-noise limited superconducting amplifiers and an optimized setup to reduce signal loss between cavities, we can efficiently track measurement-induced entanglement generation as a continuous process for single realizations of the experiment. The quantum trajectories of transmon qubits naturally split into low and high entanglement classes corresponding to half-parity collapse.  The distribution of concurrence is found at any given time and we explore the dynamics of entanglement creation in the state space. The distribution exhibits a sharp cut-off in the high concurrence limit, defining a maximal concurrence boundary. The most likely paths of the qubits' trajectories are also investigated, resulting in three probable paths, gradually projecting the system to two even subspaces and an odd subspace. We also investigate the most likely time for the individual trajectories to reach their most entangled state, and find that there are two solutions for the local maximum, corresponding to the low and high entanglement routes. The theoretical predictions show excellent agreement with the experimental entangled qubit trajectory data.
\end{abstract}
\maketitle
\date{today}

\section{Introduction}

Measurement-induced entanglement of spatially separated quantum systems is a startling prediction of quantum mechanics \cite{Hong1987,korotkov2002entanglement,ruskov2003entanglement,mao2004mesoscopic,trauzettel2006parity,williams2008entanglement, Hofer2013}. Recent experiments have demonstrated this effect via single-photon heralding \cite{Chou2005, Moehring2007, Bernien2013, Hensen2015}, as well as via continuous measurement of photons interacting with qubits \cite{roch2014observation}. The latter has the advantage of being able to investigate the physics of entanglement creation continuously, leading to new effects such as the sudden creation of entanglement after a finite measurement period, dubbed \textit{entanglement genesis} \cite{williams2008entanglement}. However, many questions are outstanding, such as (a) What is the complete characterization of the dynamics of entanglement creation as a continuous trajectory?, (b) What is the statistical distribution of the
entanglement at any time during the process?, and (c) What is the most likely way entanglement is generated?  In this work, we give a systematic answer to these questions and more, by analyzing
experimentally entangled quantum trajectories of jointly measured transmon qubits and showing excellent agreement with the theory developed here.

The rapid development of quantum information science in the superconducting domain \cite{devoret2013superconducting} has seen an exponential increase in qubit coherence time within the past decade, leading to many scientific advances \cite{siddiqi2016}. This technological progress has led to a wide variety of advances in quantum physics as observed and controlled in these systems, including $>99\%$ fidelity in single qubit quantum gates \cite{ghosh2013high}, multi-qubit entanglement generating gates \cite{fedorov2011implementation},  the violation of Bell's inequality \cite{ansmann2009violation} and quantum process tomography \cite{bialczak2010quantum}.  Recent developments include the observation of quantum states of light in resonators \cite{vlastakis2013deterministically}, as well as nearly quantum-limited parametric amplifiers \cite{castellanos2008amplification,hatridge2011dispersive}.

The improvement of coherent quantum hardware has brought with it a renewed focus on the physics of quantum measurement.  Generalized measurements have been carried out in superconducting qubits \cite{katz2006coherent}, realizing probabilistic measurement reversal \cite{korotkov2006undoing,katz2008reversal}, weak values \cite{groen2013partial,campagne2014observing}, and their connection with generalized Leggett-Garg inequalities \cite{williams2008weak}. Continuous measurements \cite{wiseman2009quantum} in superconducting systems have only recently been realized, owing to the challenge associated with high-fidelity detection of microwave signals near the single-photon level. In particular, experimental achievements include continuous feedback control \cite{vijay2012stabilizing,de2014reversing}, the tracking of trajectories in individual experiments in both the plain measurement case \cite{murch2013observing,jordan2013quantum,ibarcq2015,naghiloo2015fluores}, as well as with a concurrent Rabi drive \cite{weber2014mapping}. These experiments show detailed quantitative agreement with theory, indicating good understanding of quantities such as the most likely path of the quantum state between boundary conditions, predicted with an action principle of an associated stochastic path integral \cite{chantasri2013action,chantasri2015stochastic,jordan2015fluores}.

Going beyond a single qubit opens the possibility of measurement-induced entanglement, using a dissipative process as a tool to generate quantum correlations.  For quantum architectures building upon transitions at optical frequencies \cite{Chou2005, Moehring2007, Bernien2013, Hensen2015}, this feat has typically been achieved by relying on the correlated detection of photons at the output of a beam-splitter \cite{Hong1987} to herald an entangled state.  While powerful, this measurement protocol is binary and instantaneous, and allows no insight into the dynamical processes underlying the generation of the entangled state.  In solid state systems, such as superconducting qubit transitions in the microwave regime, there has been tremendous interest in \textit{continuously} generating bipartite \cite{ruskov2003entanglement,mao2004mesoscopic,trauzettel2006parity,williams2008entanglement} and multi-partite \cite{korotkov2002entanglement, helmer2009multiqubit,lalumi2010multiqubit} entangled states, using weak measurements that slowly interact with the qubits, in such a way that enables the resolution of the dynamical aspects of the entangling backaction. Analog feedback control can be naturally applied in the weak continuous measurement regimes \cite{wang2005entfeed,hill2008entfeed,meyer2014entfeed} and digital feedback-generated entanglement has already been demonstrated \cite{riste2013deterministic}. Joint measurement is uniquely useful as a means to generate entanglement between remote qubits \cite{roch2014observation,motzoi2015,martin2015remote,govenius2015remote,silveri2015theory, Narla2016}, for which no local coupling exists and therefore no unitary means of generating entanglement are available. 

The chief advantage of the continuous approach is in the efficiency of entanglement generation. In contrast to photon-counting schemes (in which entanglement generation rates are heavily limited by photon losses), continuous measurement enables an entanglement generation rate limited only by the pre-measurement initial state and the postselection criterion, which are both experimental choices. However, continuous measurements may be highly sensitive to dissipative losses and inefficiencies in amplification. Understanding and studying the dynamical processes underlying continuous measurement-induced entanglement is therefore critical for balancing these tradeoffs.

In this work, we combine an efficient quantum amplifier with a continuous half-parity measurement, to conduct detailed experimental and theoretical investigations of the statistics of individual trajectories of qubit pairs as they undergo the entangling process.  By peering into the ensemble, we can understand the full spectrum of evolution paths as the two-qubit state gradually projects onto the entangled subspace or onto a trivial separable state.   We explore the probability distribution of the qubits' concurrence to understand how the distribution changes in time, from a separable state with zero concurrence, to projected states in either a separable subspace or an entangled subspace. This can also be seen from the most likely path analysis, showing the emergence of different most probable paths for each final state. Moreover, we investigate the distribution of the time-to-maximum-concurrence, finding that the most probable time to maximum values of concurrence has a bimodal structure.  Studying the statistics of a large set of trajectories - rather than averaging over all of them to study the dynamics of the ensemble - enables an unprecedented understanding of the dynamics of entanglement creation under measurement.

The paper is organized as follows. In Section~\ref{ctraj}, we describe our transmon qubit measurement setup and the method of reconstructing the joint trajectories. In Section~\ref{conctraj}, we derive the concurrence-readout relation and compute the distribution of concurrence as the measurement back-action proceeds.  This distribution is then compared with the data generated from the experiment. In Section~\ref{mlp}, we turn to the most likely path analysis, finding three possible paths that the joint system takes from a separable initial state to final entangled or separable states. The experimental most likely paths are generated independently to compare with the theoretical prediction. We also discuss the distribution of the time for the two qubits to reach their maximum entanglement. The conclusions are presented in Section~\ref{conc}. Additional details of the most likely path calculations and a discussion of the parity meter are presented in the Appendices.

\section{Trajectories of transmon qubits}
\label{ctraj}

\begin{figure*}
\includegraphics[width=12.5 cm]{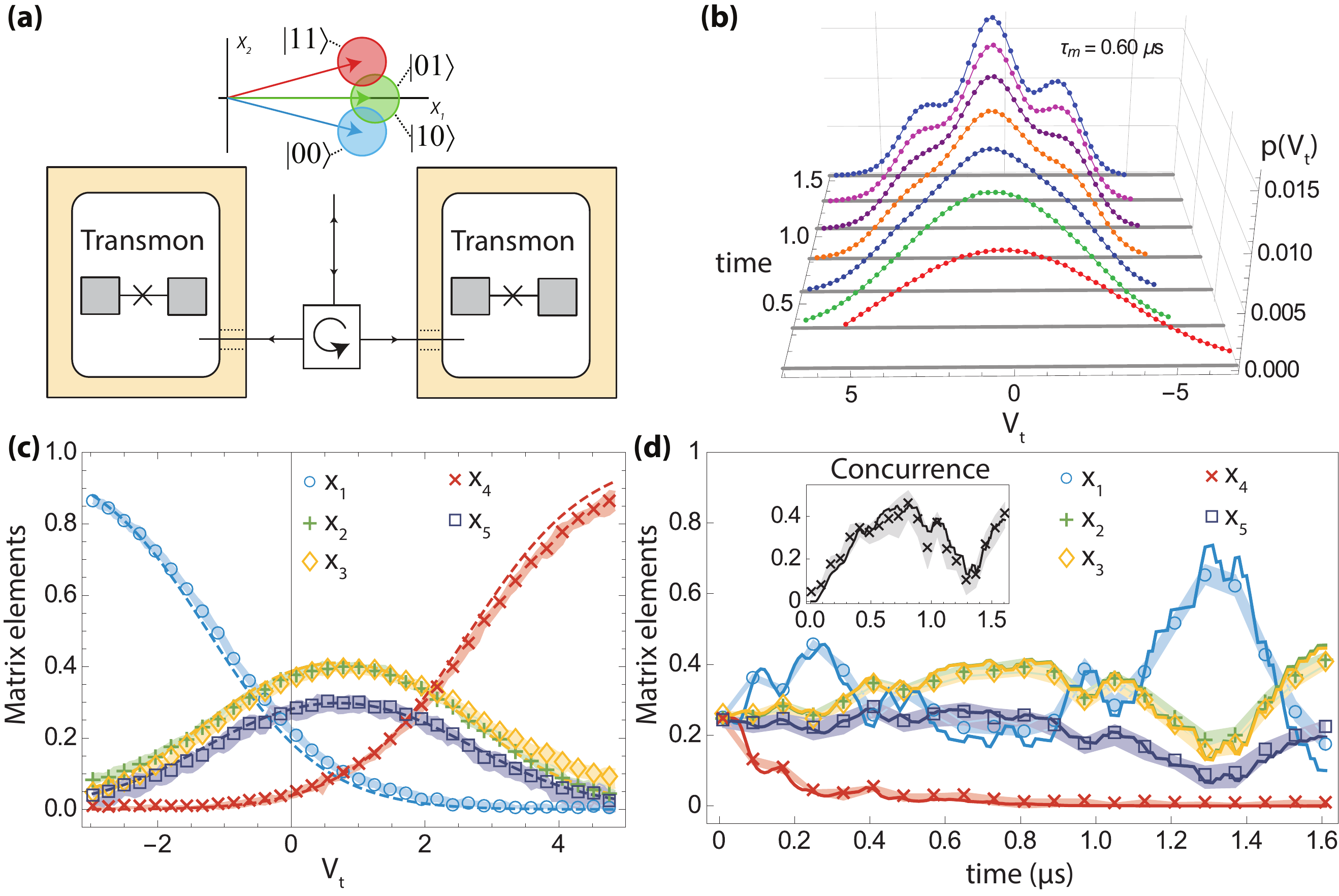}
\caption{Experimental setup and transmon qubits trajectories. (a)  shows a simplified illustration of the experimental setup. Two Transmon qubits in two remote cavities are linked via a single circulator in a \textit{bounce-bounce} geometry; the output of the circulator is routed to a high-efficiency amplification chain (not shown). The upper inset represents the distribution of the microwave field in the ($X_1,X_2$) quadrature plane at the output of the amplifier. The three possible outcomes correspond to three different subspaces: $|00\ra$, $|11\ra$, and the odd-parity subspace (spanned by the $|01\ra$ and $|10\ra$ states). (b) shows the evolution in time of the measurement outcome probability distribution $p(V_t)$. (c) shows the agreement between the experimentally-generated conditional tomography $x_i(V_t)$ (symbols and shaded error bars) and the theoretical Bayesian reconstruction (dashed lines), for a single time $t=0.48$ $\mu s$.  In panel (d), we use such reconstructions to predict and verify the trajectory of a single iteration of the experiment, showing both density matrix elements $x_i$ and concurrence (inset).}
\label{fig-setup}
\end{figure*}

We consider a two-qubit system realized by superconducting transmon qubits embedded in spatially separated microwave cavities in a setup optimized to reduce losses between the cavities. These qubits are jointly measured via a dispersive readout in a \textit{bounce-bounce} geometry (see figure~\ref{fig-setup}(a)) in which a microwave tone is sequentially reflected off of two copper cavities each containing a transmon qubit, and subsequently is amplified and measured via homodyne detection.  The cavities are directly joined by a circulator that enforces the unidirectional transfer of the coherent state. A single transmon gives a dispersive phase shift of $\pm \phi$ for states $|0\ra, |1\ra$. The shift $\phi$ is given in general by $\phi = \arg{\left[\alpha_{\ket{0}}\right]}-\arg{\left[\alpha_{\ket{1}}\right]}$, where $\alpha_i$ represents the intra-cavity coherent state conditioned on qubit state $i$. In the fully symmetric, weak measurement case (when all qubit and cavity parameters are identical and $\chi \ll \kappa$), this phase shift is given by $\phi \approx 2\chi/\kappa$, where $\chi$ is the cross-Kerr nonlinearity between the qubit and the cavity, and $\kappa$ is the cavity damping rate.

In the bounce-bounce geometry, there generally exists a probe frequency at which the dispersive shift from both transmon qubits are the same ($\phi_1 = \phi_2 = \varphi$), such that the measurement tone can acquire a phase shift of either $2\varphi, 0, 0, -2\varphi$ for states $|00\ra, |01\ra, |10\ra, |11\ra$. For small $\chi/\kappa$, this results in a \textit{half-parity} measurement on the two qubits, where the measurement result can distinguish between three subspaces: the $|00\ra$ state, the $|11\ra$ state, and the odd-parity subspace, but not within the odd parity subspace, spanned by the $|01\ra$ and $|10\ra$ states. Conditioning on the measurement results with a zero phase shift can lead to creation of an entangled state within the odd parity subspace (superpositions of $|01\ra$ and $|10\ra$).  

When the output of the second cavity is directed to a nearly-quantum limited amplification chain, the instantaneous homodyne detection signal can be correlated with the measurement back-action on the qubits, and therefore can be used to track the evolution of the system in time. By reducing the amplitude of the coherent state used to measure the system, we can engineer an entangling measurement with a characteristic measurement time ranging from several hundred nanoseconds to several microseconds: critically, these timescales are easily resolvable experimentally. The dynamics or the trajectory of the system state can be obtained via the full master equations \cite{motzoi2015,martin2015remote}, using a two-cavity polaron transformation to account for the cavity degree of freedom, giving the stochastic master equation for the qubit trajectories. Alternatively, in a limit of large cavity decay rate $\kappa \gg |\chi|$, the qubits evolution can be continuously tracked via the quantum Bayesian approach \cite{korotkov2002entanglement,ruskov2003entanglement}, inferring the current states of the system from the measurement readouts and how likely they are to occur. The two approaches both show good agreement in tracking the qubit pair state \cite{roch2014observation}.

In this paper, we focus on the quantum Bayesian approach, as it is directly related to the probability distribution of the measurement readout, and naturally leads to the probability distribution of quantum trajectories. Let us denote $p(V_t|i)$ as a probability density function of a measurement readout $V_t$ conditioned on the two-qubit states $i$, where $i = 1,2,3,4$ representing the states $|00\ra, |01\ra, |10\ra, |11\ra$ respectively. The quantum Bayesian update for this type of double-qubit measurement provides a convenient way to calculate the joint state at time $t$, given a known state at the initial time and the readout $V_t$,
\begin{align} \label{eq-bayes}
\rho_{ij}(t) = \frac{ \rho_{ij}(0) \sqrt{p(V_t| \, i ) p(V_t|\, j)} e^{-\gamma_{ij} t} }{\sum_{k=1}^4 \rho_{kk}(0) p(V_t|\, k)},
\end{align}
where $\rho_{ij}$ denotes the $ij$ element of the two-qubit density matrix and $\gamma_{ij}$ is a decoherence rate associated to the matrix element.

We define the readout  $V_t \equiv (f/t) \int_{0}^{t} {\tilde V}(t') {\rm d} t' - v_0$ as a time-average of a raw homodyne voltage signal rescaled with a weight factor $f$ and an off-set $v_0$, where $f$ is chosen so that the variance $\sigma^2_{V_t} = 1/4 \eta_m t$ is a function of a quantum efficiency of the homodyne measurement, $\eta_m  \approx 0.22$. The total probability distribution $p(V_t)=\sum_{k=1}^4 \rho_{kk}(0) p(V_t|\, k)$, shown in Figure \ref{fig-setup}(b), slowly resolves into the three peaks expected for a half-parity measurement.

The conditional readout distributions are well-approximated by Gaussian functions, giving $p(V_t|i) = (t/ \pi s)^{-1/2} \exp\{- (V_t - \delta v_i)^2t/s\}$ with the centering signals $\delta v_i = v_i-v_0$ for $i = 1,...,4$, where $s=1/2 \eta_m$. The measurement process cannot distinguish the two states in the odd-parity subspace, therefore the readout distributions corresponding to the states $|01\ra$ and $|10\ra$ are completely (or nearly) overlapped, giving $\delta v_2 \approx \delta v_3 \approx 0$ and $- \delta v_1 \approx  \delta v_4 \approx \delta v$. The measurement strength is characterized by an inverse of a characteristic measurement time $\tau_m \approx 1/\delta v^2 \eta_m$. The dephasing rates $\gamma_{ij}$ for $\delta v_i \ne \delta v_j$ are dominated by the effect of the distinguishability between states $i$ and $j$, $\gamma_{ij} \sim (\eta_m^{-1} -1) (\delta v_i-\delta v_j)^2/4s$ \cite{korotkov2002entanglement}, resulting in the strong suppression of all off-diagonal elements except $\rho_{23}$. In an ideal half-parity measurement, the decay of $\rho_{23}$ would be limited only by the intrinsic lifetimes of the qubits; however, we must additionally account for experimental imperfections in the matching of $\delta v_2$ and $\delta v_3$ and for the loss of photons between the two cavities. These effects are included in the (slightly time-dependent) dephasing rate $\gamma_{23}$.

Since we expect most of the off-diagonal terms to damp quickly, we only consider five density matrix elements: $x_1 \equiv \rho_{11}$, $x_2 \equiv \rho_{22}$, $x_3 = \rho_{33} $, $x_4 \equiv \rho_{44} $, and $ x_5 \equiv |\rho_{23}|$  \footnote{Theoretically, if there is no single qubit unitary rotation and the initial matrix elements are real values, all elements would stay real during the measurement process. However, in the experiment, the two cavities are not exactly at the same frequency and the qubits are not equally coupled to their respective cavities, so the qubits experience different AC-stark shifts. The off-diagonal element acquires a deterministic (i.e. predictable) non-zero phase that increases linearly in time and, critically, does not affect the degree of entanglement in the system.  We therefore study the magnitude of this element, and neglect the phase.}.  In order to compare the Bayesian prediction to the true density matrix, we perform conditional tomography \cite{chow2010entexp, roch2014observation} in order to experimentally reconstruct the full experimental mapping $V_t \mapsto \rho(V_t, t)$.  In Figure \ref{fig-setup}(c), we show good agreement between the Bayesian prediction and conditional tomography reconstruction of $x_i(V_t)$ for a single measurement time $t = 0.48$ $\mu s$.  We show examples of transmon qubit trajectories in Figure \ref{fig-setup}(d) for an initial state prepared in a product of ${\hat x}$-states such that $x^0_1=x^0_2=x^0_3=x^0_4=x^0_5=1/4$.  We show both the Bayesian reconstruction and the tomographic verification, which are in good agreement.  This verifies that the Bayesian reconstruction can be used to faithfully translate a noisy measurement signal into the stochastic evolution of a joint quantum state.

\section{Concurrence trajectories and their distribution}\label{conctraj}
As a measure of entanglement between two parties such as the transmon qubits, concurrence is a convenient choice and can be computed directly from the density matrix of the system \cite{wooters1998}. The concurrence formula for the half-parity setup is greatly simplified because of the suppression of most matrix elements, resulting in a X-shape density matrix, of which the concurrence is calculated from \cite{wooters1998,jakobczyk2005},
\begin{align}\label{eq-conc0}
{\cal C}(t) = 2\,\text{max}\left\{ 0, x_5(t)- \sqrt{x_1(t) x_4(t)} \right\}.
\end{align}
The value of concurrence ranges from $0$ for a separable state to $1$ for the Bell states. The concurrence trajectory of an exemplar trajectory is shown in the inset to Figure \ref{fig-setup}(d). In this section, we will use the simplified formula Eq.~\eqref{eq-conc0} to show that the concurrence of the two-qubit state can be determined directly from the measurement readout, which then leads to the derivation of the concurrence probability distribution as a function of the measurement readout and measuring time.

\subsection{Concurrence-readout relationship}
\label{cv}
Let us consider the concurrence formula in Eq.~\eqref{eq-conc0}, where its value is determined by the second term in the bracket, which we denote $c_t \equiv 2 \left\{ x_5(t)- \sqrt{x_1(t) x_4(t)} \right\}$. If $c_t$ is a non-negative quantity (i.e., $c_t\ge 0$), then the concurrence is simply given by ${\cal C}(t) = c_t$. We will show at the end of this section that this is always the case for our chosen initial qubit state and parameter regimes, but is not true in general \cite{williams2008entanglement}. From the Bayesian update in Eq.~\eqref{eq-bayes}, we calculate the quantity,
\begin{align} \label{eq-conc1}
c_t = \,\frac{2}{{ N}}\left\{ x_5^0 \sqrt{ { P}_2 {P}_3}e^{-\gamma\, t} - \sqrt{x_1^0x_4^0 \,{ P}_1 { P}_4 } \right\},
\end{align}
where $x_i^0$ for $i = 1,...,5$ are the matrix elements of the initial qubits state, and $\gamma = \gamma_{23} $. We have used simplified notations ${P}_i  \equiv p(V_t|i)$ for the probability distributions, and $N$ for a normalized factor given by ${ N} =\sum_{k = 1}^{4} x_k(0) P_k$.

Substituting the probability distribution functions with the Gaussian functions of the means $\delta v_i$ for $i = 1,...,4$, we obtain a form of $c_t$ explicitly as a function of $V_t$ and $t$,
\begin{align}\label{eq-conc2}
c_t(V_t,t) =\frac{2}{\cal M} \left\{ x_5^0 e^{({ \alpha}_{23} V_t - { \beta}_{23}-\gamma)t} - \sqrt{x_1^0 x_4^0}e^{({ \alpha}_{14} V_t - { \beta}_{14})t} \right \} ,
\end{align}
where the prefactor is given by ${\cal M} = \sum_{i=1}^4 x_i^0  e^{2 \alpha_iV_t  t- 2 \beta_i t}$ using a set of defined variables: $\alpha_i = \delta v_i/s$, ${ \alpha}_{ij} = \alpha_i+\alpha_j$, $\beta_i = \delta v_i^2/2 s$, ${ \beta}_{ij} = \beta_i+\beta_j$, and $s=1/2 \eta_m$.

The quantity $c_t$ in Eq.~\eqref{eq-conc2} would represent the actual concurrence of the qubits state at any time $t$, if $c(t)\ge 0$ is satisfied. For our chosen initial state, a product of single qubit ${\hat x}$-states: $x_1^0 = x_2^0 = x_3^0 =x_4^0 = x_5^0 = 1/4$, the quantity $c_t$ is non-negative whenever a condition $(\gamma - { \alpha}_{23} V_t + { \beta}_{23}) <  ( { \beta}_{14}-{ \alpha}_{14} V_t)$ is true. From the experimental data (e.g., for the setup with $\tau_m = 0.60 \mu s$), we have $-\delta v_1 \approx \delta v_4$ and $\delta v_2 \approx \delta v_3 \approx 0$ (giving ${ \alpha_{14}} , \alpha_{23}, \beta_{23} \approx 0$) and ${ \beta_{14}} \sim 3.2$ MHz, while $\gamma < 0.6$ MHz. Therefore, the condition is always satisfied and the second term in the bracket of Eq.~(\ref{eq-conc2}) decays faster than the first term, resulting in an always non-negative quantity. Consequently, the quantity in Eq.~\eqref{eq-conc2} gives the concurrence-readout relationship ${\cal C}(V_t,t) = c_t(V_t,t)$, and the concurrence at any time $t$ can be determined directly from the time-averaged measurement readout $V_t$.

The concurrence formula in Eq.~\eqref{eq-conc2} can be simplified further by considering a perfectly symmetric half-parity measurement, $\delta v_2 = \delta v_3 = 0$ and $-\delta v_1 = \delta v_4 = \delta v$. Given the initial state, a product of two qubit ${\hat x}$-states, the concurrence is then given by,
\begin{align}\label{eq-conc3}
{\cal C}_{\text{ps},{\hat x}}(V_t,t) = \frac{e^{-\gamma t} - e^{-\delta v^2 t/s}}{ 1 + \cosh(2 \,V_t \, \delta v \, t/s) e^{-\delta v^2 t/s}},
\end{align}
where the subscript `ps,${\hat x}$' indicates the perfectly symmetric half-parity measurement given the specific initial state.

\subsection{Probability density function for concurrence trajectories}
From the direct relationship between the measurement readout and the concurrence of the qubits state, the probability density function of the concurrence can be derived from the probability distribution of the time-averaged signal $V_t$. The distribution of the time-averaged readout is given by,
\begin{align}
p(V_t  ) = \sum_{i=1}^4 x_i^0 \, p(V_t|i)\label{eq-probv}.
\end{align}
The variance of the distribution $\sigma^2_{V_t} = s/2 t$ narrows as time increases, leading to the collapse of the joint qubit state into three categories: $|00\rangle$ state, $|11 \rangle$ state, and some superposition state of $|01\rangle$ and $|10\rangle$ after a few characteristic measurement times $\tau_m$.

\begin{figure}
\includegraphics[width=8cm]{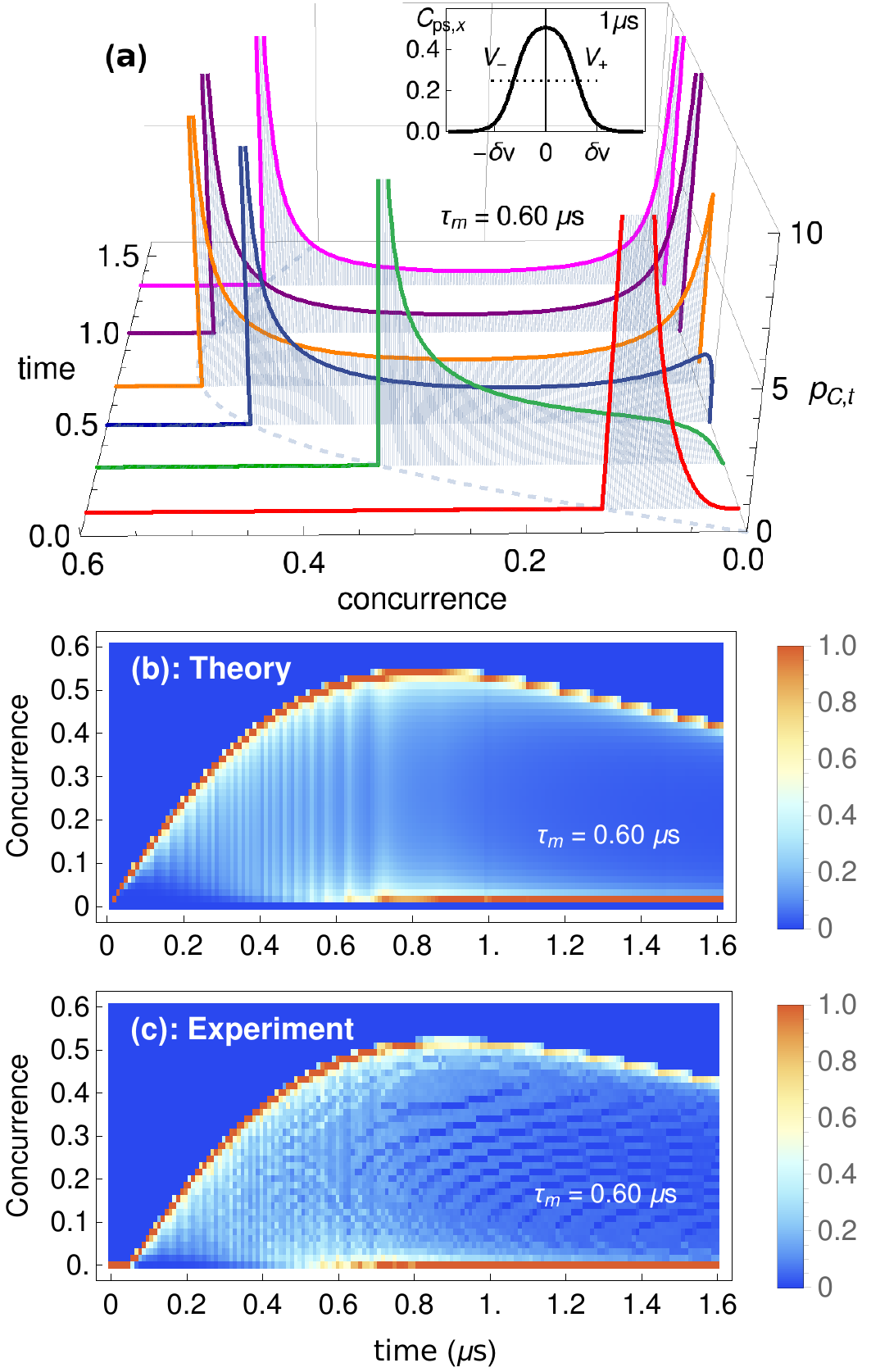}
\caption{Concurrence distribution for the qubits under the half-parity measurement. In the panel (a), we plot the concurrence probability density function Eq.~\eqref{eq-probc} for different values of time. The values of time for the presented curves are chosen so as to see their unique features as they develop. The grey dotted curve joining the high-concurrence peaks shows the concurrence upper bound Eq.~\eqref{eq-maxconc}. The inset shows an example of how the concurrence (at time $t = 1 \mu s$) varies as a function of the readout $V_t$. (b) and (c) are the histograms of the concurrence at any time points from $t=0$ to $T=1.6 \mu s$ (with a step size $0.01 \mu s$), comparing theory and experimental data. For the theory plot, we coarse-grain the distribution $p_{{\cal C},t}(c)$ in Eq.~\eqref{eq-probc} by integrating it with a pixel size $\delta c \sim 0.015$, which is the bin size of the experimental histogram. For presentation purposes, a histogram at any time $t$ is normalized by its maximum element.}
\label{fig-conc}
\end{figure}

Knowing the probability density function of the time-averaged signal, we follow the transformation of random variables $V_t \rightarrow {\cal C}$ using the concurrence-readout relationship Eq.~\eqref{eq-conc2} (or \eqref{eq-conc3} for a perfectly symmetric case). The concurrence ${\cal C}(V_t,t)$ is not a monotonic function in $V_t$; instead it has a bell-like shape as shown in the inset of Figure~\ref{fig-conc}(a). We write the cumulative distribution function of the concurrence $F_{{\cal C},t}(c)  = p_{{\cal C},t}({\cal C} \le c) = p( V_t \le V_-) + \{1- p(V_t \le V_+)\}$, where $p_{{\cal C},t}(c)$ is a probability density function for the concurrence, and $V_+$, $V_-$ are two solutions that arise from solving Eq.~\eqref{eq-conc2} (or \eqref{eq-conc3}), ${\cal C}(V_t, t) = c$. The concurrence distribution is then obtained by taking a derivative of the cumulative distribution,
\begin{align}  \label{eq-probc}
p_{{\cal C},t}(c) =p(V_-)\bigg|\frac{\partial V_-}{\partial c} \bigg| + p(V_+)\bigg|\frac{\partial V_+}{\partial c}\bigg|,
\end{align} 
noting that $V_-(c,t)$ and $V_+(c,t)$ are functions of the concurrence $c$ and time $t$. The full solution of $p_{{\cal C},t}(c)$ is quite lengthy and is not shown.

We show in Figure~\ref{fig-conc}(a) the plots of concurrence probability distributions \eqref{eq-probc} for different values of time $t$, and in Figure~\ref{fig-conc}(b,c) the density plot comparing between the theory and the transmons experiment. At an early time, the distribution of concurrence is narrowly peaked near its maximum which increases over time, whereas at later times, a second peak emerges near the zero concurrence, showing a bimodal distribution. In Figure~\ref{fig-conc}(b), the theoretical histogram for the concurrence is obtained by integrating the theory curves Eq.~\eqref{eq-probc} for the probability over small intervals $\delta c \approx 0.015$. This is to make a fair comparison with the histogram of the experimental data in Figure~\ref{fig-conc}(c), calculated with a bin size of $0.015$. We note that a short delay in the experimental entanglement creation is a result of the cavity ring-up time, which will be discussed more in the next section.

We stress that the concurrence distribution has a sharp upper bound (shown as a grey dotted curve in Figure~\ref{fig-conc}(a)), which the concurrence cannot exceed. In order to understand why the probability distribution for the concurrence has a sharp upper cut-off at any time, we recall that the density matrix of the two-qubit system, conditioned on the time-integrated readout $V_t$, is entirely specified by that (random) outcome, together with the initial state, the dephasing rate, and other parameters of the problem, Eq.~\eqref{eq-bayes}. Consequently, the concurrence is controlled by $V_t$, as in Eq.~\eqref{eq-conc3}.  As can be seen from the inset of Figure~\ref{fig-conc}(a), the concurrence, plotted as a function of the measured signal $V_t$ is bounded from above for any fixed time $t$ by some amount we call ${\cal C}_{\rm max}$, and consequently, any value of concurrence higher than that maximum (whose value will change as the time increases) cannot be realized.  Therefore the probability distribution of concurrence has a sharp upper cut-off given by ${\cal C}_{\rm max}(t)$. Physically, this indicates that there is an upper limit on how fast entanglement can be created by the continuous measurement in this situation, even for rare events of the measurement process.  

For the perfectly symmetric case in Eq.~\eqref{eq-conc3}, we can find an analytic solution for the upper bound of the concurrence, knowing that $\cosh(x)$ has its minimum at $x=0$, ${\cal C}_{{\rm ps},{\hat x}}(V_t)$ has its maximum at $V_t =0$, so consequently, the concurrence upper bound is given by 
\begin{align}\label{eq-maxconc}
{\cal C}_{\text{max},{\rm ps},{\hat x}}(t) = \frac{e^{-\gamma t} -  e^{-\delta v^2 t/s}}{ 1+ e^{-\delta v^2 t/s}}.
\end{align}
The behaviour of this bound is a result of two competing rates, between the extra dephasing rate $\gamma$ and a measurement rate $\delta v^2/s$. Eq.~\eqref{eq-maxconc} increases from zero for small time and decays for long time after reaching its maximum concurrence as seen in Figure~\ref{fig-conc}. The maximum possible concurrence for this qubit half-parity measurement and the time this happens can be obtained from this relation. More about the time to reach maximum concurrence will be discussed in Section~\ref{sec-timedist}.

\section{Most likely path analysis}
\label{mlp}
In addition to the distribution analysis in the previous section, where we treated the concurrence at each point in time independent from any other times, we now incorporate the notion of connected trajectories, finding a probability density function for quantum trajectories and their most likely paths. These most likely paths describe routes with highest probability density, taking into account that each of the points in the trajectory ensemble are connected via quantum state update rules, e.g. in Eq.~\eqref{eq-bayes}. As we have seen previously, the concurrence distribution exhibits a transition from a single-peak distribution to double-peak distributions, one at the upper bound concurrence and another near zero concurrence. Here, with the notion of trajectories, we will also see that the two-qubit state starting in its initial state gradually collapses to three subspaces: the $|00\ra$ subspace, the $|11\ra$ subspace and the odd subspace, as described by three most likely routes.

\begin{figure*}
\includegraphics[width=18cm]{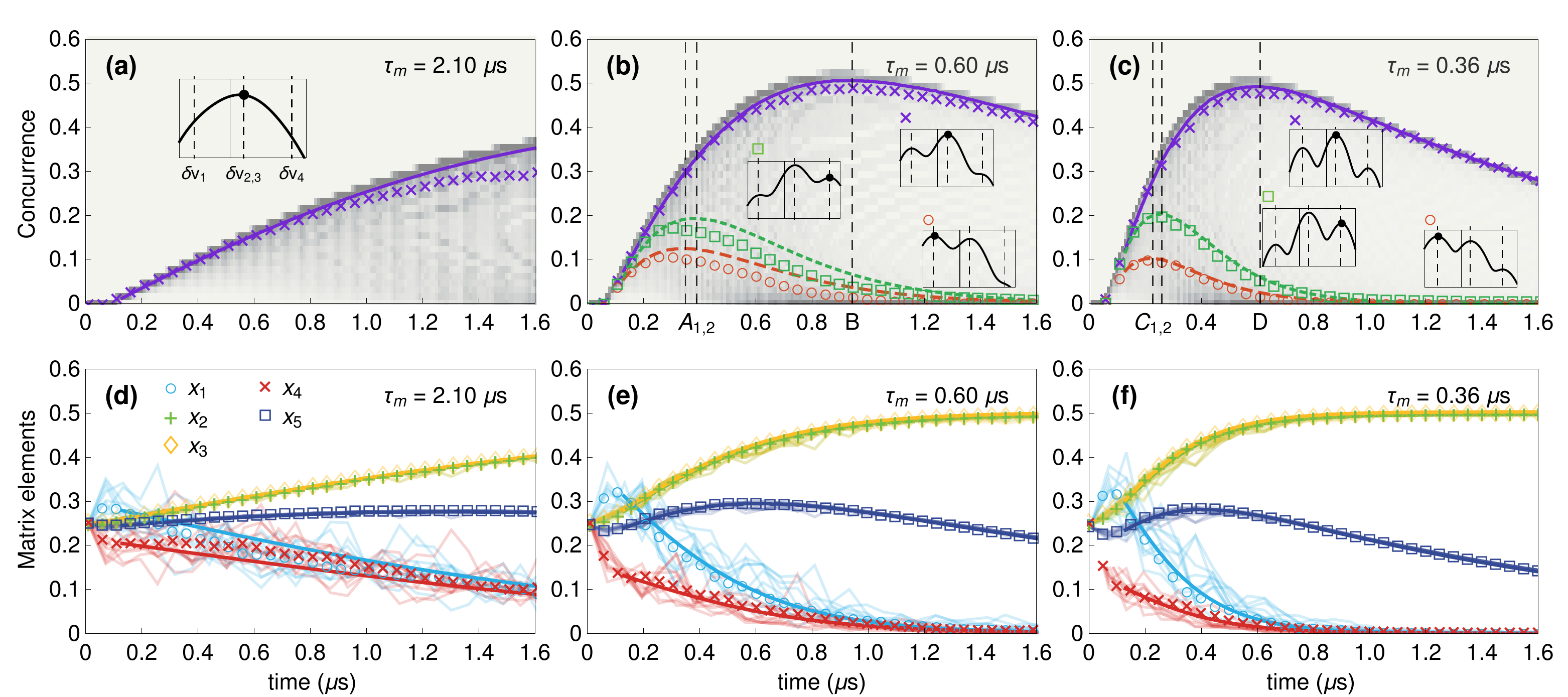}
\caption{The most likely paths from the theoretical prediction and the experimental data. The first, second, and third columns are from three different data sets, measuring the same qubits and cavities with three different measurement readout powers (indicated by the characteristic measurement time): $\tau_m = 2.10 \mu s, \, 0.60 \mu s, \, 0.36 \mu s$, respectively. The top row panels (a,b,c) show the concurrence of the multiple most likely paths: the path with high concurrence (solid magenta), and the two paths with low concurrence projecting onto $|00\ra$ (dashed orange) and $|11\ra$ (dotted green) subspaces. The vertical lines (labelled as $A_{1,2}, B, C_{1,2}, D$) represent the times at which the concurrence is maximum for each of these paths. The insets show the log-likelihood as functions of optimal measurement readouts, and the greyscale histograms in the background shows experimental concurrence histograms similar to the one in Fig.~\ref{fig-conc}(c). The bottom row panels (d,e,f) present the evolution of the density matrix elements of the high-concurrence most likely paths: the solid curves indicating theoretical solutions, whereas the data points indicating experimental most likely paths postselected with the most entangled state at the final time $T = 1.6 \mu s$. Examples of the post-selected trajectories are shown as fluctuating curves.}
\label{fig-mlp}
\end{figure*}

Let us consider the five non-trivial elements of the two-qubit density matrix $\{ x_1, x_2, x_3, x_4, x_5\}$, and treat each element as an independent variable. Each state trajectory is represented by a time series of these five elements, which corresponds to one realization of the measurement readout $\{v_t \} = \{ v_{0}, v_{\delta t}, v_{2 \delta t}, ..., v_{t}\}$ where we define, $v_t = (f/\delta t)\int_t^{t+\delta t}{\tilde V}(t') {\rm d} t' - v_0$, as an instantaneous readout at time $t$ with an integration time $\delta t$. Since the readout $v_t$'s are assumed Markovian and only depend on the qubit states right before the measurement, a joint probability density function for the readout realization is given by,
\begin{align}\label{eq-likelihood}
{\cal P}(\{ v_t\}) = \prod_{t'=0}^t \left\{ \sum_{k=1}^4 x_{k,{t'}}\, p(v_{t'}|k) \right\},
\end{align}
a product of probability distributions of $v_{t'}$ from $t'=0$ to $t'=t$ with a time step $\delta t$. The probability function for an instantaneous readout is given by $p(v_t|i) = \sqrt{\delta t/\pi s}\, \exp\{-(v_t - \delta v_i)^2 \delta t/s\}$ for $i = 1,2,3,4$.

To derive the most likely path for these two-qubit trajectories, we use the Bayesian update equation for the two-qubit state Eq.~\eqref{eq-bayes}, adapted to a state update every time step $\delta t$, and then we maximize the joint probability density Eq.~\eqref{eq-likelihood} constraining the state update equations. Following the most likely path analysis for quantum states under continuous measurement \cite{chantasri2013action,chantasri2015stochastic} and introducing Lagrange multipliers $\{p_1, p_2, p_3, p_4, p_5\}$ for the contraints, we obtain differential equations for an optimal path in the qubits state space,
\begin{subequations}\label{eq-master}
\begin{align}
\partial_t x_i = & + \frac{x_i}{s} \sum_{k=1}^4 x_{k}\big\{ 2 v_t ( \delta v_i-  \delta v_k) - \delta v_i^2  + \delta v_k^2 \big\},\\
\nonumber\partial_t x_5= &-\gamma x_5 + \frac{x_{5}}{s}\bigg\{v_t(\delta v_2+\delta v_3)-\frac{(\delta v_2^2+\delta v_3^3)}{2} \\
&+ \sum_{k=1}^4 x_{k}( \delta v_k^2 - 2 v_t \delta v_k)\bigg\},
\end{align}
\end{subequations}
where $i = 1,...,4$ for the first line and the variables $x_k$ are time-dependent functions. We note that $v_t$ in these equations behaves as a ``smooth'' optimal readout, which is a function of the qubit density matrices and their Lagrange multipliers, determining an optimal path from an initial state to its final state \cite{chantasri2013action}. An optimal path is a solution of 10 differential equations: 5 for the qubit state variables Eq.~\eqref{eq-master}, another 5 for Lagrange multipliers, and the optimal readout as a function of both sets of variables (see Appendix~\ref{app-mlp}). These equations describe most likely paths for a measurement with any values of $\delta v_{1,...,4}$ and can also be generalized to include the effect of external drives on the two-qubit. However, in the absence of external drive, these can be simplified as we will see in the next section.

\subsection{Most likely paths for joint measurement of transmon qubits}
In this work, where the transmon qubits only evolve with the influence of the measurement back-action, the optimal readout is found to be constant in time (see Appendix~\ref{app-mlp}). We can therefore bypass solving the full set of the differential equations \cite{silveri2015theory}, and only compute the qubit evolution in Eq.~\eqref{eq-master} with constant $v_t$, looking for measurement results with maximum likelihood. To estimate the likelihood of a two-qubit trajectory, we evaluate a logarithm of the probability density function Eq.~\eqref{eq-likelihood} approximated to first order in $\delta t$ \cite{chantasri2013action} to obtain,
\begin{align}\label{eq-logprob}
\log{{\cal P}(\{ v_t \})} \approx {\cal S}_0 - \!\! \int_0^t\!\!\! {\rm d} t' \, \left\{\frac{1}{s}\sum_{k=1}^4 (v_{t'} - \delta v_k)^2 x_{k,t'}\right\},
\end{align}
where ${\cal S}_0$ represents a state-independent part of the joint probability function ${\cal P}$. We show in the insets of Figure~\ref{fig-mlp}(a,b,c) examples of the approximated log-likelihood as functions of optimal readout $v_t$, for three different measurement strengths. In the case of $\tau_m \approx 2.10 \mu s$, there is only one maximum likelihood value located at $v_t \approx \delta v_{2,3}$, which gives the most likely path with high concurrence shown as a solid curve in Figure~\ref{fig-mlp}(a); whereas, in the stronger measurement cases, Figure~\ref{fig-mlp}(b) and (c), the approximated log-likelihood have three local maxima: the middle ones, $v_t \approx \delta v_{2,3}$, corresponding to most likely paths collapsed to entangled states (high-concurrence branches), and the sided peaks, $v_t \approx \delta v_1$ and $v_t \approx \delta v_4$, corresponding to two branches of most likely paths collapsed to $|00\ra$ and $|11\ra$ states (low-concurrence branches), respectively. We note that this technique of finding multiple most likely paths is not under final-state constraints as in \cite{chantasri2013action,weber2014mapping}.

In order to test the theoretical most likely paths, we need to extract the most likely paths from the experimental data. We collect an ensemble of $10^4$ transmon trajectories and then compute average trace distances between any two trajectories, (e.g., $\rho_a$ and $\rho_b$),
\begin{align}
D_{a,b} = \frac{\delta t}{2 T} \sum_{t'=0}^{t } \Tr{\sqrt{(\rho_a(t')- \rho_b(t'))^{\dagger}(\rho_a(t')- \rho_b(t'))}},
\end{align}
for all possible pairs. The goal is to pick the first few trajectories with minimum total distance to other trajectories in the ensemble, and average them to get an estimate of the \textit{experimental most likely paths}. For this particular set of data, we choose $\sim 10^2$ highly likely trajectories to get a smooth estimate of the most likely paths. However, for the case that there exist multiple (e.g., three) most likely paths, we first divide the ensemble into subensembles according to their trace distance, and then apply the minimum total distance procedure to the trajectories in each subensemble separately. In Figure~\ref{fig-mlp}(a,b,c), we show the concurrence of the experimental most likely paths as data points.

Ideally, the theoretical most likely paths predicted from the log-likelihood Eq.~\eqref{eq-logprob} for each set of measurement strength using the initial state $ \{ x_1^0,...,x_5^0\} = \{ 1/4, ..., 1/4\}$ would be good enough to compare with the experimental data. However, in the experiment, after the initial state has been prepared, the cavities take some time to reach their steady state condition, making the parameters $\delta v_1, ..., \delta v_4$ unstable during the first $\sim 0.13 \mu s$ time. Therefore, we need to let the initial qubit state evolve and then find new ``initial'' states at time $t = 0.13 \mu s$ for the theoretical most likely path calculation. We use the experimental most likely states at $t=0.13 \mu s$ (one for each branch) as the initial states in calculating the log-likelihood functions (the insets), which then lead to an excellent prediction of the most likely paths and their concurrence for the rest of the evolution. 

We also note that the unequal two low-concurrence branches in Figure~\ref{fig-mlp}(b,c) happen because the population of the states drifts more toward the ground state $|00\ra$ during the transient time, as a result of the qubit relaxation during the measurement. Moreover, in Figure~\ref{fig-mlp}(d,e,f), we present the evolution of matrix elements of the high-concurrence most likely paths, showing a good agreement between the theoretical most likely paths and experimental most likely paths post-selected with the most entangled state at the final time $T= 1.6 \mu s$.

\subsection{Most likely paths for perfectly symmetric half-parity measurement}
We are also interested in finding analytic solutions for the most likely paths for the symmetric case: $\delta v_2 = \delta v_3 = 0$ and $-\delta v_1 = \delta v_4 = \delta v$. The differential equations \eqref{eq-master} for the qubit state simplify to,
\begin{subequations}\label{eq-simplify}
\begin{align}
\partial_t x_p =&- b x_p   -a x_m + a x_p x_m + b x_p^2, \\
\partial_t x_m = & - b x_m - a x_p  + b x_m x_p + a x_m^2,\\
\partial_t x_{2,3} = & +x_{2,3} (a x_m  + b x_p),\\
\partial_t x_5 = & - \gamma x_5 + x_5( a x_m + b x_p),
\end{align}
\end{subequations}
where we have defined new variables $x_p \equiv x_1 + x_4$ and $x_m \equiv x_1 - x_4$, and constant parameters $a = v_t \delta v/s$ and $b = \delta v^2/4s$. The first two equations can be solved independently from the rest. For the case when $v_t = \delta v_{2,3}= 0$, which corresponds to the most likely odd-parity result, we obtain an analytic solution for the high-concurrence branch of the most likely path,
\begin{subequations}
\begin{align}
x_{1,4}(t) \propto& \,\, x_{1,4}^0 \exp(-\delta v^2 t/4 s),\\
x_{2,3}(t) \propto& \,\, x_{2,3}^0 ,\\
x_5(t) \propto &\,\, x_5^0 \exp(-\gamma t),
\end{align}
\end{subequations}
where the proportionality factor is the inverse of a normalized factor ${\cal N} = (1 - x^0_1 - x_4^0) + (x_1^0 + x_4^0)\exp(-\delta v^2 t/4 s)$. The concurrence of this path exactly coincides with the upper concurrence bound derived in Eq.~\eqref{eq-maxconc}, which is not surprising because the distribution of concurrence shows sharp peaks along the concurrence bound. We note that the solutions for the low-concurrence branch (for $v_t =  \delta v$ and $v_t = - \delta v$) can be found numerically. 

In the above analysis, for the half-parity case, a simple analytic solution for the most likely path for arbitrary values of $v_t$ was not forthcoming. However, if we further simplify the problem by considering a parity meter \cite{williams2008entanglement}, then we can solve the equations of motion, Eqs.~\eqref{eq-master} and their conjugate equations, exactly. A parity meter has the same detector outputs for the even and odd parity subspaces, but the detector can distinguish between the subspaces. More detail of the calculation is presented in Appendix~\ref{app-fullparity}.

\subsection{Distribution of time to maximum concurrence}\label{sec-timedist}
In the process of the entanglement generation, there are interesting quantities to investigate such as the maximum concurrence each individual trajectory can reach, and the time it takes to reach the highest value. We previously showed that qubit trajectories branch out to high and low concurrence subspaces. Therefore, one would expect that there are \textit{two} most likely times for the qubit trajectories to reach their maximum concurrences (or their most entangled states).

\begin{figure}
\includegraphics[width=8.6cm]{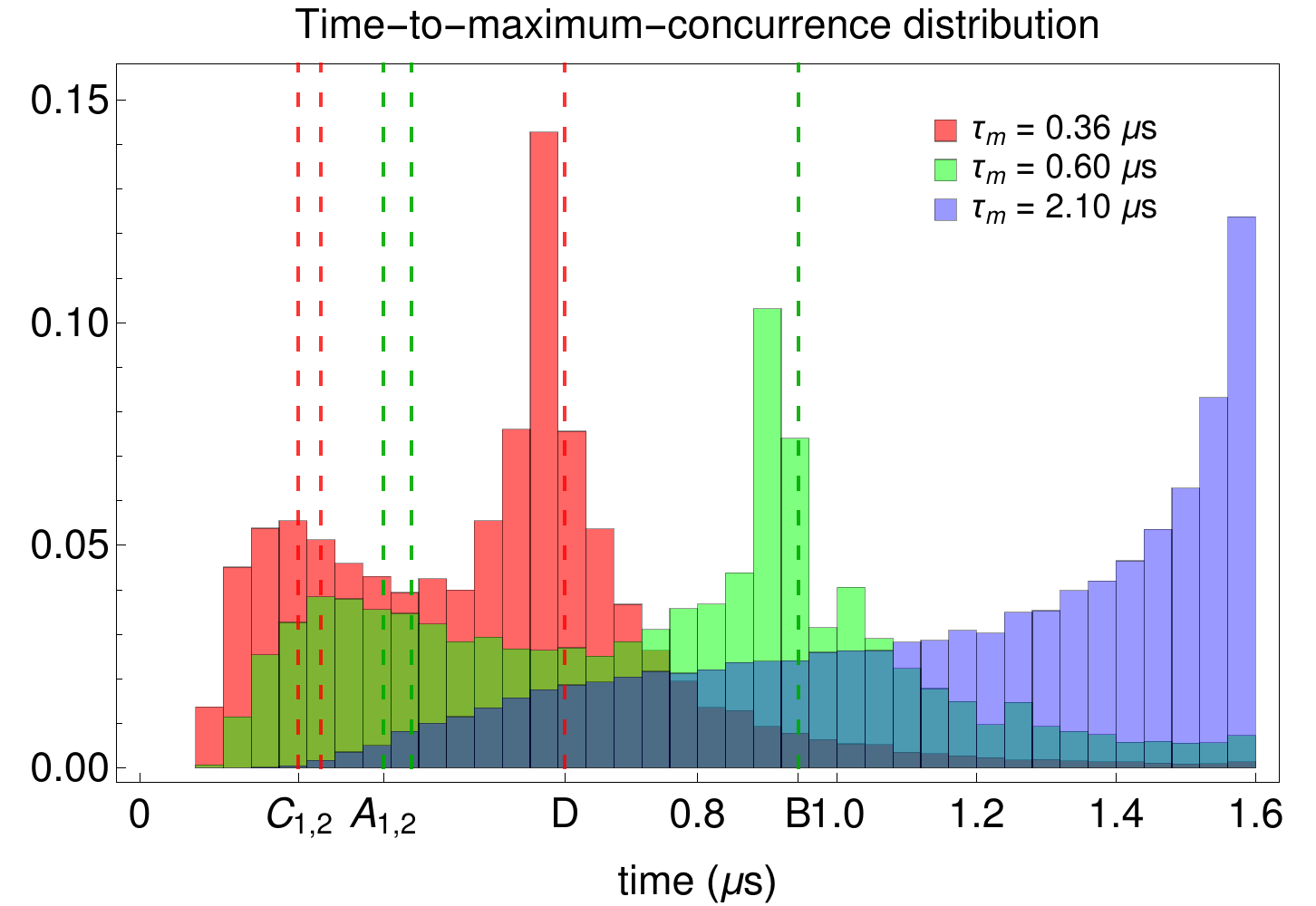}
\caption{Histograms of the time-to-maximum concurrence for individual trajectories for three measurement strengths indicated by the values of $\tau_m$. For the two cases with strong readout powers (shown in red and green histograms), there exists two peaks corresponding to two most likely times to reach their most entangled states. The theoretical prediction of these times are shown as vertical dashed lines labelled as $A_{1,2}, B, C_{1,2}, D$ (same as in Figure~\ref{fig-mlp}(a,b,c)).}
\label{fig-timedist}
\end{figure}

We show in Figure~\ref{fig-timedist} the normalized histograms of time for transmon qubit trajectories to reach their maximum concurrence. The histograms for the $\tau_m = 0.36 \mu s$ and $0.60 \mu s$ measurement cases explicitly show double peaks, which agree with the branching of concurrence and the most likely qubit paths in Figure~\ref{fig-mlp}(b,c). The times at which these peaks are located can be theoretically predicted from the time-to-maximum-concurrence of the solutions of the most likely paths; as shown by the vertical dashed lines in Figure~\ref{fig-mlp}(b,c) and Figure~\ref{fig-timedist}: $A_{1,2}$, $B$ are the two most likely times to reach maximum concurrence (for low and high concurrence branches, respectively) for $\tau_m = 0.60 \mu s$, and $C_{1,2}$, $D$ are the same but for the case with $\tau_m = 0.36 \mu s$. The agreement between the theoretical prediction of the peaks and the peaks of the histograms are as good as the agreement of the theory-experiment most likely paths in Figure~\ref{fig-mlp}(b,c). We note that for the weak measurement regime, $\tau_m = 2.10 \mu s$, the bifurcation has not occurred yet during the measurement time $T = 1.6 \mu s$. One would expect to see a branching effect, when the total measurement time is long enough.

\section{Conclusion}
\label{conc}
We have investigated the process of entanglement generation between two spatially separated superconducting transmon qubits, and their statistical properties. The entanglement of the two qubits is created as a result of the half-parity dispersive measurement, via the microwave pulses sequentially interacting with both qubits. The strength of the joint measurement is arbitrary and we have studied three different values of the measurement strength. We examined the concurrence of individual trajectories and theoretically calculated its distribution from the quantum Bayesian approach, gradually projecting the two-qubit states to entangled states with high concurrence, and to separable states with zero concurrence.

The most likely path analysis was also carried out, predicting the most probable paths for the qubits trajectories. We found that in the two-qubit state space, there are three likely paths conforming to the three branches projecting to the $|00\ra$ subspace, the $|11\ra$ subspace, and the odd ($|01\ra, |10\ra$) subspaces; the first two correspond to the lower branches of the concurrence bifurcation, and the last corresponds to the high concurrence branch. These theoretical most likely paths show excellent agreement with the experimental most likely ones extracted from the transmon trajectories data (for three independent data sets) based on the trace distance between trajectories in two-qubit state space. Moreover, we have presented the distributions of the time to the maximum concurrence for individual trajectories. The most likely path analysis was shown to be useful in predicting the peaks of these time distributions.

We conclude that the accurate tracking of quantum trajectories of a jointly measurement qubit system is possible, and that the physics of the entanglement creation statistics is well described by a quantum trajectory theoretical approach.  The theoretical most likely paths to entanglement and concurrence distribution match the experiment excellently. This work shows the way to use this process as a control mechanism to entangle remote systems for quantum information processing purposes.  In future work, similar questions can be posed about the full parity measurement, and we have made some predictions about that case in this work.

\begin{acknowledgments}
This work was supported by US Army Research Office Grants No. W911NF-15-1-0496 and No. W911NF-13-1-040, by National Science Foundation grant DMR-1506081, by John Templeton Foundation grant ID 58558, and by Development and Promotion of Science and Technology Talents Project Thailand. M.E.K.S. acknowledges support from the Fannie and John Hertz Foundation.
\end{acknowledgments}

\appendix

\section{The optimized paths (most likely paths) with pre-/post-selected states}\label{app-mlp}

Following the outline in Ref.~\cite{chantasri2013action,chantasri2015stochastic}, for the action principle for the continuous quantum measurement, the optimized paths starting from an initial state and ending at a final state after some time $t$ is given by optimizing an action of the stochastic path integral,
\begin{align}\label{eq-action}
{\cal S} = - \int_0^t  {\rm d}t' \bigg\{  \frac{1}{s} \sum_{i=1}^4 (v_{t'} - \delta v_i)^2 x_{i}+ \sum_{j=1}^5  p_j (\partial_t x_j - {\cal F}_j)  \bigg\},
\end{align}
where the two-qubit variables $x_j$ and their conjugates $p_j$ for $j=1,2,...,5$ are implicitly functions of time $t'$. The first sum in the time integral is the logarithm of the joint probability density function of the measurement readout ${\cal P}(\{v_t\})$ truncated to first order in $dt$, and the functional ${\cal F}_j$ is the right hand side of Eq.~\eqref{eq-master}.

By extremizing the action Eq.~\eqref{eq-action} over all variables $x_i, p_i,v_t$, we get a set of 10 ordinary differential equations (ODEs) for the optimized path, and one equation for optimal measurement readout. The set of ODEs includes 5 differential equations of the two-qubit variables $x_i$'s, as shown in Eq.~\eqref{eq-master}, and 5 equations for the conjugate variables $p_i$'s,
\begin{subequations}\label{eq-pmaster}
\begin{align}
\partial_t p_i =& \sum_{j=1}^4 x_j A_{ij} + x_5 p_5 B_{i} + C_i \quad \text{for $i=1,2,3,4$},\\
\partial_t p_5 = &+\gamma_5 p_5 + p_5\sum_{j=1}^4 x_j B_{j},
\end{align}
\end{subequations}
where,
\begin{align}\nonumber
A_{ij} =& -  (p_i-p_j)(2 v_t - \delta v_i-\delta v_j)(\delta v_i-\delta v_j),\\ \nonumber
B_{i} = &\frac{1}{2s}\bigg\{v_t(4 \delta v_i - 2 \delta v_2- 2 \delta v_3)-(2 \delta v_i^2-\delta v_2^2- \delta v_3^2)\bigg\},\\ \nonumber
C_i = & \frac{1}{2s}(v_t-\delta v_i)^2.
\end{align}
The optimal readout is given as a function of the two-qubit variables and the conjugate variables,
\begin{align}\label{eq-v}
 \nonumber v_t =& \sum_{i=1}^4 \sum_{j=1}^4 \bigg\{p_i x_i   x_j (\delta v_i -\delta v_j)  + x_i \delta v_i  \bigg\} \\
&+ \frac{p_5 x_5}{2} \sum_{i=1}^4(\delta v_2 +\delta v_3-2 x_i \delta v_i ).
\end{align}
connecting the ODEs for the qubit variables and the conjugate variables.

By taking time-derivative of the function in Eq.~\eqref{eq-v}, and substituting both $\partial_t x_i$ and $\partial_t p_j$ with the ODEs above, we find that $\partial_t v_t = 0$.

\section{Analytic solution in the case of a parity meter}\label{app-fullparity}

For the full-parity measurement case, we assume $\delta v_1 = \delta v_4 =d v$, while $\delta v_2=\delta v_3=0$. Putting in this special case, we find the equations of motion,
\begin{subequations}\label{eq-ana}
\begin{align}
\partial_t x_1 =&+ \lambda\, x_1 x_o, \\
\partial_t x_2 =& - \lambda\, x_2 x_e, \\
\partial_t x_3=& - \lambda\, x_3 x_e, \\
\partial_t x_4 =& +\lambda\, x_4 x_o, \\
\partial_t x_5=& -\gamma x_5 - \lambda x_5 x_e,
\end{align}
\end{subequations}
where we define $\lambda = 2 v_t  \delta v/s - \delta v^2/s$, and $x_e = x_1 + x_4 = \rho_{00,00} + \rho_{11,11}$ is the probability of being in the even parity subspace, while $x_o = x_2+x_3 = \rho_{01,01}+\rho_{10,10}$ is the probability of being in the odd parity subspace.

Taking the sum of ${\partial_t x_1}$ and ${\partial_t x_4}$, we can derive an equation for $x_o$ alone since $x_e + x_o=1$,
\begin{align}
{\dot x_o} = - {\dot x_e} = -\lambda (1-x_o) x_o.
\end{align}
Integrating this equation gives the solution,
\begin{align}
x_o(t) = \frac{x_{o}^0 e^{-\lambda t}}{1 - x_{o}^0 (1-e^{-\lambda t} )},
\end{align}
where $x_{o}^0$ is the initial condition for $x_2+x_3$.  Similarly, we find for the even probability,
\begin{align}
x_e(t) = \frac{x_{e}^0}{1 - x_{o}^0 (1-e^{-\lambda t})},
\end{align}
where $x_{e}^0$ is the initial condition for $x_1+x_4$.

Notice that the value of $\lambda$ may be found completely from these results.  The parity probability of the initial pre-selected and final post-selected state will fix the value of $\lambda$. If the integrated signal gives a positive answer larger than $d v/2$, $\lambda$ will be positive, tending to collapse the state into the even parity subspace.  Conversely, if the integrated signal is less than $d v/2$, then $\lambda$ will be negative, tending to collapse the system into the odd parity subspace.

From these solutions, we may find the other density matrix solutions.  For example, the Eq.~(\ref{eq-ana}a) may be rewritten as ${\partial_t\ln x_1} = \lambda \,x_o$, which can be integrated to get $x_1(t) = x_{1}^0 \exp \big\{\lambda\int_0^t {\rm d} t'  x_o(t')\big\} $. We simply apply this method to the rest of Eqs.~\eqref{eq-ana} and obtain the solutions, the most likely path of the two-qubit problem,
\begin{subequations}
\begin{align}
x_1(t) =& \frac{x_{1}^0}{1 - (x_{2}^0+x_{3}^0) (1-e^{-\lambda t} )},\\
x_2(t) = &\frac{x_{2}^0 e^{-\lambda t}}{1 - (x_{2}^0+x_{3}^0) (1-e^{-\lambda t} )},\\
x_3(t) = &\frac{x_{3}^0 e^{-\lambda t}}{1 - (x_{2}^0+x_{3}^0) (1-e^{-\lambda t} )},\\
x_4(t)  = &\frac{x_{4}^0}{1 - (x_{2}^0+x_{3}^0) (1-e^{-\lambda t} )},\\
x_5(t) =& \frac{x_{5}^0 e^{-(\gamma+\lambda) t}}{1 - (x_{2}^0+x_{3}^0) (1-e^{-\lambda t} )},
\end{align}
\end{subequations}
where the value of $\lambda$ can be fixed once we have specified the initial and final values of the density matrix, $x_{1}^0,x_{2}^0,...,x_{5}^0, x_{1,f}, x_{2,f},...,x_{5,f}$.

Finally, we may find the concurrence of the most likely path by calculating ${\cal C} = 2\,{\rm max} \big\{0, x_5 - \sqrt{x_1 x_4}\big\}$. We find the result,
\begin{align}
{\cal C}(t) =2\, {\rm max} \bigg\{0, \frac{x_{5}^0 |e^{-(\gamma + \lambda) t}| - \sqrt{x_{1}^0 x_{4}^0}}{ {|1 - (x_{2}^0+x_{3}^0) (1-e^{-\lambda t} )}|}\bigg\}.
\end{align}
In order to cross the entanglement border \cite{williams2008entanglement} from an unentangled state, we must have $\lambda$ take on a negative value to enhance the first term over the second, taking the system into the odd parity subspace.

The entanglement border will be crossed when the concurrence is zero, or when the condition
\begin{align}
\gamma + \lambda = (1/t) \ln\bigg( \frac{x_{5}^0}{\sqrt{|x_{1}^0 x_{4}^0|}}\bigg),
\end{align}
is satisfied.

%

\end{document}